\documentclass[12pt,a4paper]{article}
 \usepackage{graphicx}
 \usepackage{afterpage}
 \usepackage{enumerate}
  \usepackage{longtable}
\usepackage{latexsym}
\usepackage{amsfonts}

\begin{document}
\small
 \title{\bf The influence of the medium physical conditions and atomic constants on the Stokes profiles of absorption lines in the solar spectrum
}
\author{\bf V.A. Sheminova }
  \date{}

 \maketitle
 \thanks{}
\begin{center}
{ Main Astronomical Observatory, National Academy of Sciences of Ukraine,\\ Akademika  Zabolotnoho 27,  Kyiv,  03143, Ukraine,  e-mail: shem@mao.kiev.ua
}
\end{center}

\begin{abstract}
{The Stokes profiles of Fe I lines in the photosphere of the Sun are calculated within the Unno-Beckers-Landi-Degl`Innocenti theory. Estimates of the magnetic strengthening  of the lines were obtained. The changes in the Stokes profiles depending on the excitation potential, wavelength, equivalent width, Lande factor, micro- macroturbulent velocities, radial velocity, damping constant, atmospheric model, magnetic field strength and direction are considered. The graphically presented variations of the Stokes profiles make it possible to determine the initial values of the input parameters for solving the problems of magnetic field vector reconstruction by the inversion method. The presented dependencies of the magnetic strengthening on the line parameters will help to correctly select magnetically sensitive lines for the investigation of sunspots, flux tubes, plages, and other magnetic features.

}
\end{abstract}


\section{Introduction }

The results of observations and simulations of structural inhomogeneities in the solar atmosphere obtained in recent decades show a growing interest in the effects of the magnetic field in spectral lines. Particular attention is focused on the development of magnetic flux tube models, which are actively used in studies of the small-scale structure of the magnetic field. Magnetic flux tubes are characterised by strong magnetic fields and small dimensions. Strong magnetic fields polarise the emerging radiation and significantly change the profile of spectral line and increase its equivalent width.

Studies of the structure of the magnetic fields in the Sun by analysing spectral lines, which Unno \cite{13} began in 1958, continue to the present day. For example, the reconstruction of the structure of the magnetic field from observational data was carried out by Auer et al. \cite{4}, Balandin \cite{1}, Skumanich et al. \cite{9}. The influence of the gradient of macro-velocities and the magnetic field strength on the shape of the Stokes profiles was investigated by Grigoriev et al. \cite{6}.  The dependencies of the magnetic strengthening on the number of components and the distance between them, on the inclination angle  and strength  of magnetic field, on the line strength, on the microturbulent velocity and damping constant were obtained by simple calculations in the works of Boyarchuk et al. \cite{2}. Empirical line half-width dependencies on line strength, excitation potential and wavelength were also established by statistical analysis of 402 iron lines of Stenflo et al. \cite{10}. However, so far  little attention has been paid to quantitative estimates of the magnetic strengthening of the lines.

 The aim of this work is to show the influence of the input calculation parameters on the shape of the Stokes profiles of spectral lines, to quantify this influence and to reveal the main trends in the behaviour of magnetoactive lines in the presence of a magnetic field.

The paper is organised as follows.  In section 2 we present the initial data and calculation tactics, the representation of Stokes parameter profiles used and the definition of the magnitude of line strengthening in the magnetic field. Our results on the influence of the atomic line parameters and the line formation medium are presented in Sect. 3 and 4, respectively. The analysis of the anomalous dispersion effect is presented in Section 5.  Our conclusions are presented in Section 6.

\section{Calculations of Stokes profiles of spectral lines}

The Fe I line of 525.021 nm was chosen for the calculation. It has the following parameters: excitation potential $EP = 0. 12$ eV; oscillator strength $\log gf = -4.89$; Lande factor $g_{ef} = 3.0$; central depth $R = 0.714$ and equivalent width $W = 6. 35$ pm observed at the centre of the solar disk.  We adopted the LTE approximation, HOLMU solar atmosphere model \cite{7}, microturbulent velocity $V_{mic}= 0.8$ km/s, and magnetic field with with strength of $H= 1500$ G, an inclination $\gamma =55^\circ$, an azimuth of $\varphi=30^\circ$. According to the classical Unsold approximation, the damping constant due to the Van der Waals forces between the absorbing and perturbing atoms is $\gamma_{WdW}$ multiplied by the correction factor $E= 1.5$. We call the Stokes profiles of the FeI 525.021 nm line calculated with the above values of the line and medium parameters as standard. All calculations are performed  for the centre of the solar disk   ($\cos \theta=1$) with using the SPANSATM program  \cite{3}. Our assumptions are acceptable for estimating the effect of the magnetic field on photospheric lines.

The calculation tactic is as follows. To investigate the influence of the input parameter, we form sets of fictitious lines on the basis of the Fe I 525.02 line with the equal equivalent width $W$. In each set, one input parameter is modified within reasonable limits typical of spectral analysis of the solar atmosphere. The values of all other input parameters remain unchanged and correspond to the chosen standard. We then calculate the equivalent width $W$ of the lines without regard to the magnetic field, adjusting the corresponding value of the oscillator strengths so that the calculated value of $W$ for each line matches the standard. For each set of lines, we calculate Stokes profiles and equivalent line widths $W_H$ in the presence of a magnetic field.

In this paper we consider four Stokes parameters $I, ~Q,~ U,~ V$ and four Stokes profiles respectively.  The $I, ~Q,~ U,~ V$ parameters of polarised radiation are defined as follows:
\begin{eqnarray}
 I &  =&I_0+I_p=I_0 + \sqrt{Q^2+U^2+V^2},\nonumber\\
 Q &  =&I_{lin}(\delta=0^{\circ})-I_{lin}(\delta=90^{\circ}),\nonumber \\
 U &  =&I_{lin}(\delta=45^{\circ})-I_{lin}(\delta=135^{\circ}), \\
 V &  =&I_{circ}(right)-I_{circ}(left),\nonumber
\end{eqnarray}
where $ \delta $ is the angle between the vector of linearly polarized radiation and a certain direction determined by the measuring equipment. The $ I $ parameter is the total intensity of the unpolarized $ I_0 $ and polarized $ I_p $ components of the radiation. The parameters $ Q, ~ U $ represent the intensities of the linearly polarized radiation $ I_{lin} $. Moreover, $ Q $ is the difference between the intensities passed through linear polarizers, which are oriented according to the angles $ \delta = 0^\circ $ and $ \delta = 90^\circ $, and $ U $ is the same, but with orientation of polarizers $ \delta = 45^\circ $ and $ \delta = 135^\circ $. The $ V $ parameter represents the difference between the circular components of the right and left polarized radiation $ I_{circ}^{right} $, $ I_{circ}^{left} $. Right-circular polarization means that the electric field vector of the atom at a fixed point in space rotates clockwise around the magnetic field vector if the magnetic field is directed towards the observer. In the case of left-hand circular polarization, the electric vector rotates counterclockwise. It is generally accepted that the magnetic field has a positive polarity if its vector is directed towards the observer, and negative polarity if the vector is directed into the interior of the Sun. In the XYZ coordinate system with the Z axis located along the line of sight towards the observer, we determine the direction of the magnetic field by the tilt angle $ \gamma $ between the magnetic field vector and the Z axis, and the azimuth $ \varphi $, which is counted from the X axis counterclockwise in the XY plane.

The Stokes parameter profiles have been calculated according to the following expressions:
\begin{eqnarray}
R_I(\lambda)&=&(I_c-I(\lambda))/I_c=1-I(\lambda)/I_c,\nonumber\\
R_Q(\lambda)&=&(Q_c-Q(\lambda))/I_c=-Q(\lambda)/I_c,\nonumber\\
R_U(\lambda)&=&(U_c-U(\lambda))/I_c=-U(\lambda)/I_c,\\
R_V(\lambda)&=&(V_c-V(\lambda))/I_c=-V(\lambda)/I_c,\nonumber
\end{eqnarray}
where: $I_c, ~Q_c,~ U_c,~ V_c$ are the Stokes parameters of continuous radiation, which is assumed to be unpolarized, i.e.,  $ Q_c = U_c = V_c = 0$. To calculate the magnetic strengthening of the line, we used the expression:
 \begin{equation}
 q = \log (W_H / W)
 \end{equation}
 according to \cite{2}.  Here $W_H$ is the equivalent width calculated in the presence of a magnetic field and $W$ is the equivalent width of the same line without the magnetic field. Recall that the equivalent width $W$ of the lines used is the standard value of 6.36 pm. In addition, we also calculated the percentage strengthening of the line as
\begin{equation}
\Delta = (W_H-W)/W*100.
\end{equation}
Note that the magnetic strengthening depends on the number of splitting components and the distance between them, as well as on the blending processes between the components. All components are divided into three groups. The number of components is determined by the quantum numbers of the transition of a given line. The distance between components in the group is  equal
\begin{equation}
\delta = 1.4 *10^6 (g_i - g_k) \lambda H/V_{mic} ,
\end{equation}
where $g_i$ and $g_k$ are the Lande factors of the lower and upper levels. Blending between components in the group depends on the $ \delta $ distance. Blending between groups of components depends on the angle of inclination of the magnetic field vector $ \gamma $ and the distance between the centers of gravity of the groups. The increase in the equivalent width due to magnetic broadening depends mainly on the shape of the absorption coefficient. This is because the presence of a magnetic field in the absorbing medium only affects the shape of the absorption coefficient and does not change the number of absorbing particles. Magnetic broadening of weak and very strong lines is negligible or almost absent. Therefore we use the sets of moderate lines which are sensitive to the type of absorption coefficient. This will allow us to easily trace dependencies of the Stokes profiles and magnetic enhancement depend on the input parameters.

 All the results of our calculations with variations of the input parameters are presented in  the tables and figures. The tables show the following data: values of the input parameter whose influence we are studying, equivalent line width $W_H$ in the presence of magnetic field $H$, magnetic strengthening $q$ and percentage strengthening of the equivalent width $\Delta$. The figures illustrate the profiles of the four Stokes parameters $R_I$, $R_Q$, $R_U$, $R_V$, calculated for a given set of lines, and the dependence of the magnetic strengthening on the considered parameter according to the data presented in the tables.

\section{Influence of line parameters on Stokes profiles}

We first consider the effect of wavelength, excitation potential, equivalent width, Lande factor, and van der Waals attenuation constant on the Stokes profiles of lines formed in the presence of a magnetic field.

\begin{figure}
 \centerline{
\includegraphics [scale=0.8]{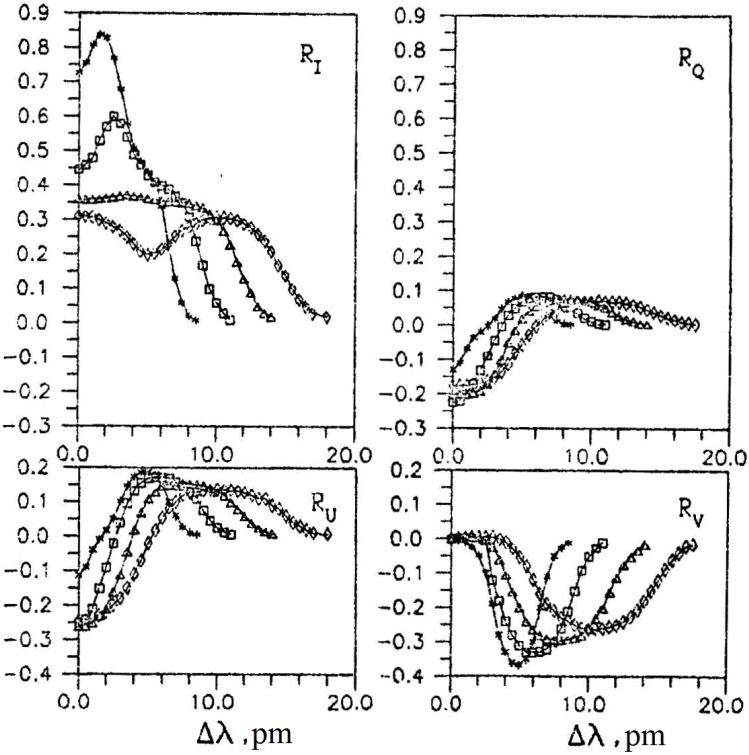}
    }
  \caption {\small\label{1f}
   Stokes profiles of the lines with wavelength $\lambda=400$ ($\ast$), 500 ($\square$), 600 ($\triangle$), 700 nm ($\lozenge$) calculated in the presence of  the magnetic field with $H=1500$~G, $\gamma=55^\circ$, and  $\varphi=30^\circ$.   }

 \centerline{
\includegraphics [scale=0.7]{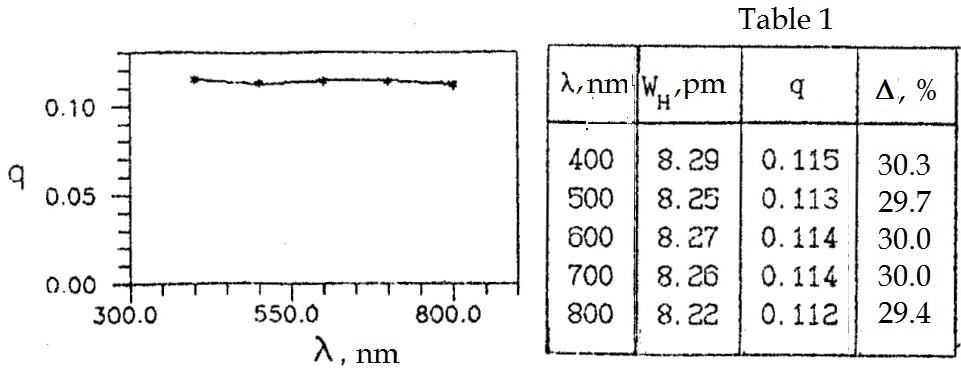}
     }
 \caption {\small\label{2f}
Dependence of the magnetic strengthening $q$   on the wavelength of the line $\lambda$.
\vspace {0.5cm}
\newline Table 1.  The magnetic strengthening $q$ and $\Delta$ of the lines  with the different $ \lambda$. }
\end{figure}

\textbf{Line wavelength.} Fig. \ref{1f} shows that the Stokes profiles become wider as the line wavelength $\lambda$ increases from 400 to 700 nm, and their extremums decrease. Despite significant changes in the shape of the profiles, the equivalent width practically remains the same and the magnetic strengthening is almost independent of the wavelength and on average $\Delta\approx 30$\% (Fig. \ref{2f}, Table 1). It follows that the wavelength of the line has no effect on the magnetic strengthening  of the line.

\begin{figure}
 \centerline{
\includegraphics [scale=0.8]{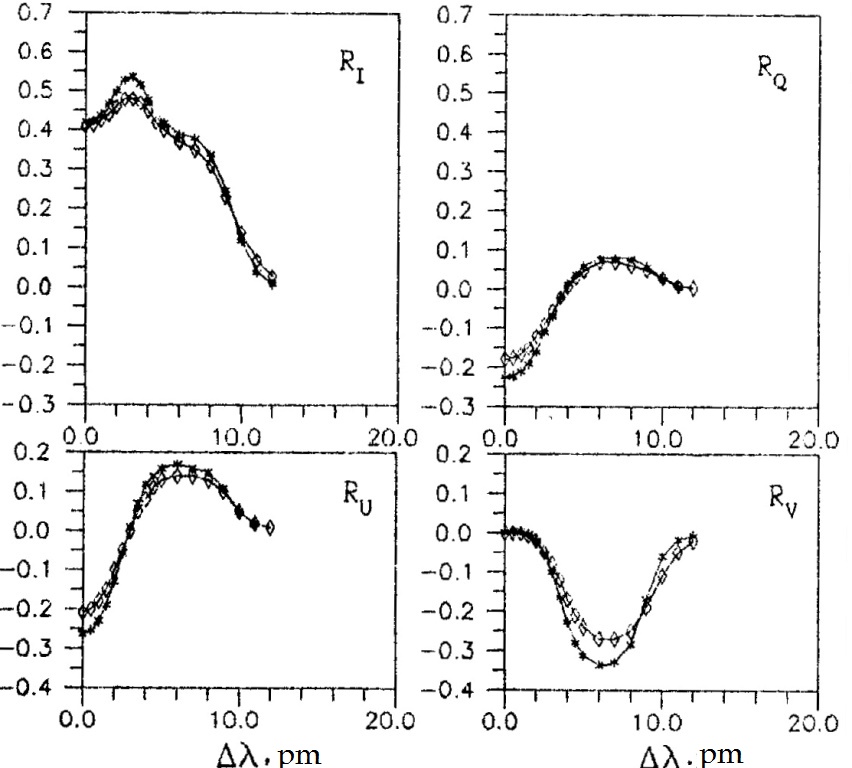}
    }
  \caption {\small\label{3f}
   Stokes profiles of the lines with the potential excitation $EP = 0$ ($\ast$) and 4 eV ($\lozenge$) calculated in the presence of  the magnetic field with $H=1500$~G, $\gamma=55^\circ$, and  $\varphi=30^\circ$.
}
 \centerline{
\includegraphics [scale=0.7]{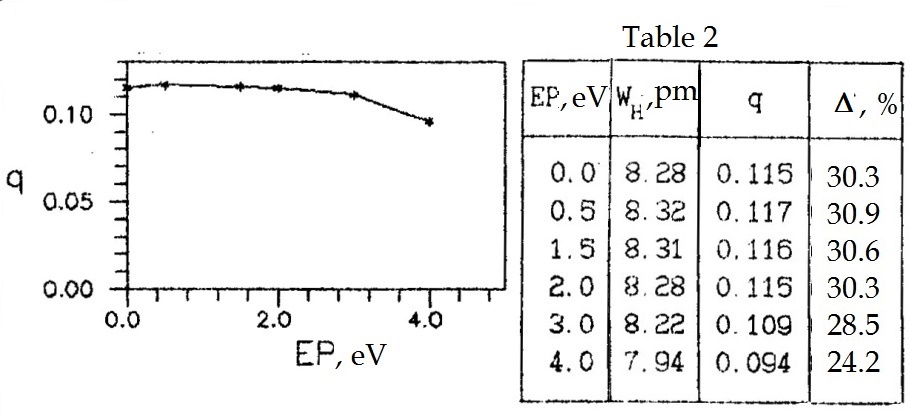}
     }
 \caption {\small\label{4f}
Dependence of the magnetic strengthening $q$  on the potential excitation of the line $EP$.
\vspace {0.5cm}
\newline Table 2.  The magnetic strengthening $q$ and $\Delta$ of the lines  with the different $EP$. }
\end{figure}

\textbf{Excitation potential.} Fig. \ref{1f} shows that the shape of the Stokes profiles varies much less with the excitation potential than with the line wavelength.  The extrema of the Stokes profiles only slightly decrease with increasing line excitation potential, and the width of the profile wing does not change.  The dependence of $q$ on $EP$ is insignificant (Fig. \ref{4f}). It can be noted that the magnetic strengthening  slightly decreases (Table 2) in the moderate Fe I lines with EP from 3 to 4 eV.

\begin{figure}
 \centerline{
\includegraphics [scale=0.8]{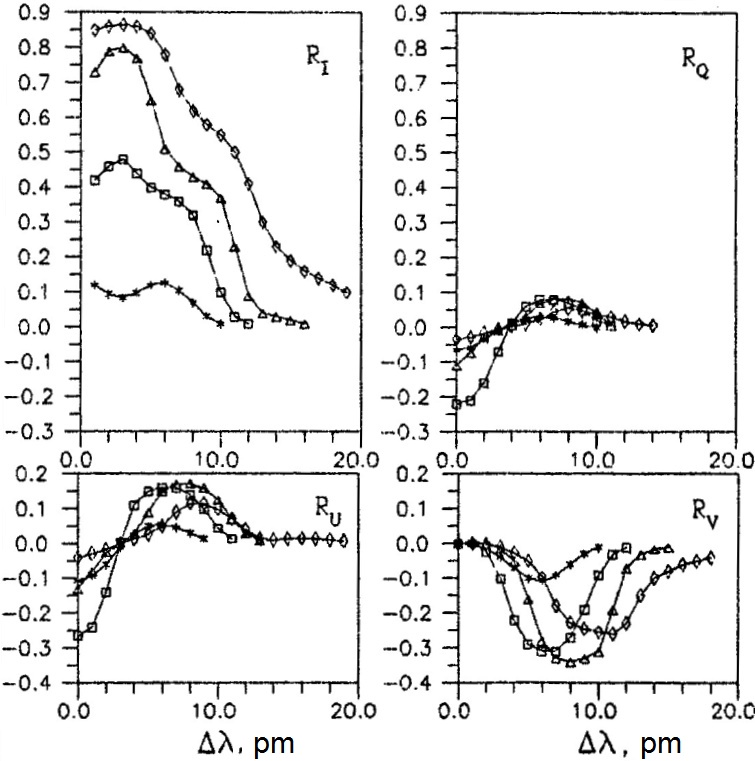}
    }
  \caption {\small\label{5f}
   Stokes profiles of the lines with the equivalent width   $W$ = 1.76 ($\ast$), 5.98($\square$), 9.95 ($\triangle$),  and 19.67 pm ($\lozenge$) calculated in the presence of  the magnetic field with $H=1500$~G, $\gamma=55^\circ$, and  $\varphi=30^\circ$.
}
 \centerline{
\includegraphics [scale=0.7]{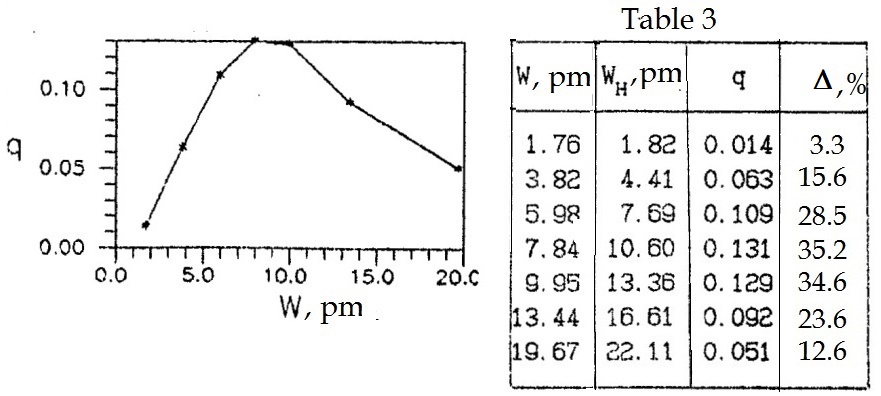}
     }
 \caption {\small\label{6f}
Dependence of the magnetic strengthening $q$ on the equivalent width  $W$.
\vspace {0.5cm}
\newline Table 3.  The magnetic strengthening $q$ and $\Delta$ of the lines  with the different $W$. }
\end{figure}

\textbf{Equivalent width or line strength.} Here we used a set of lines with different $W$ from 2 to 20 pm for the calculations.  Quite significant changes in the Stokes profiles occur depending on the line strength or equivalent width $W$. The greatest changes in the shape of the $R_I$ profile can be seen in weak and moderate lines (Figure \ref{5f}). The magnetic strengthening is greatest in lines with $W= 7$--11 pm, and smallest in the weakest lines (Fig. \ref{6f} and Table 3).  In Table 3 we have also presented the  equivalent widths $W$ of the lines used.  Our calculations confirm the fact that moderate and moderately strong lines are most sensitive to the magnetic field.


 \begin{figure}
 \centerline{
\includegraphics [scale=0.8]{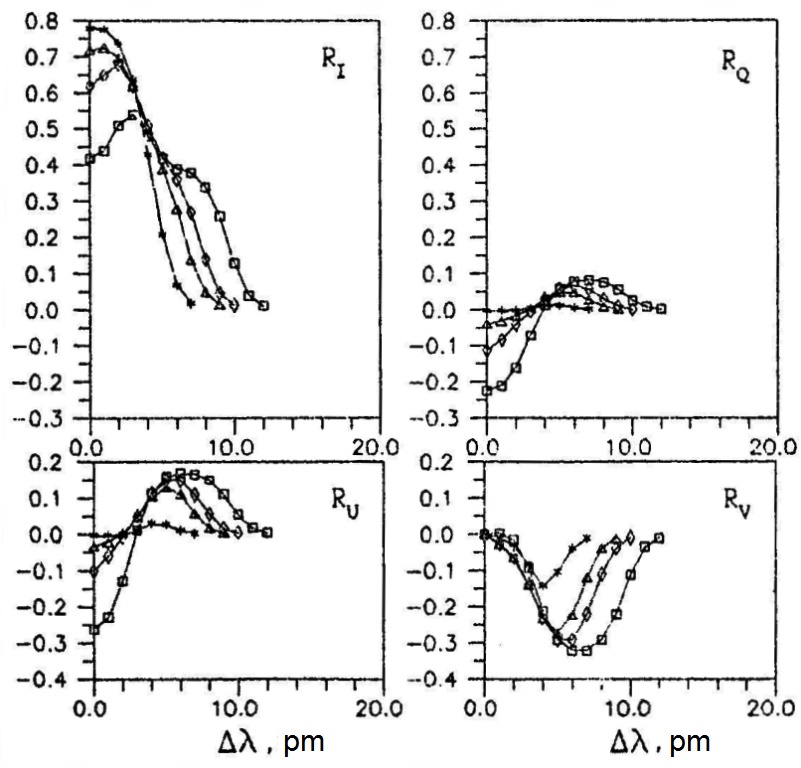}
    }
  \caption {\small\label{7f}
   Stokes profiles of the lines with the Lande factor $g_{ef} =0.5$   ($\ast$), 1.5 ($\triangle$),  and 2 ($\lozenge$), and 3 ($\square$) calculated in the presence of  the magnetic field with $H=1500$~G, $\gamma=55^\circ$, and  $\varphi=30^\circ$.
}
 \centerline{
\includegraphics [scale=0.7]{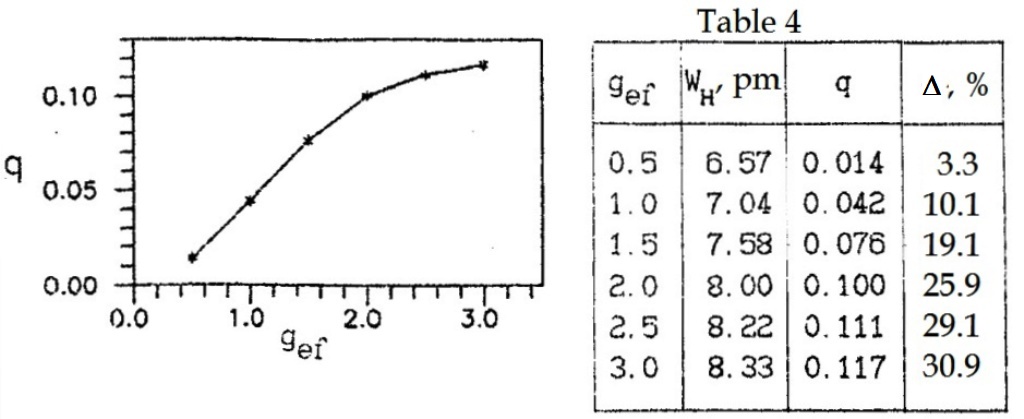}
     }
 \caption {\small\label{8f}
Dependence of the magnetic strengthening $q$ on the Lande factor $g_{ef}$.
\vspace {0.5cm}
\newline Table 4.  The magnetic strengthening $q$ and $\Delta$ of the lines  with the different $g_{ef}$. }
\end{figure}
 \begin{figure}
 \centerline{
\includegraphics [scale=0.8]{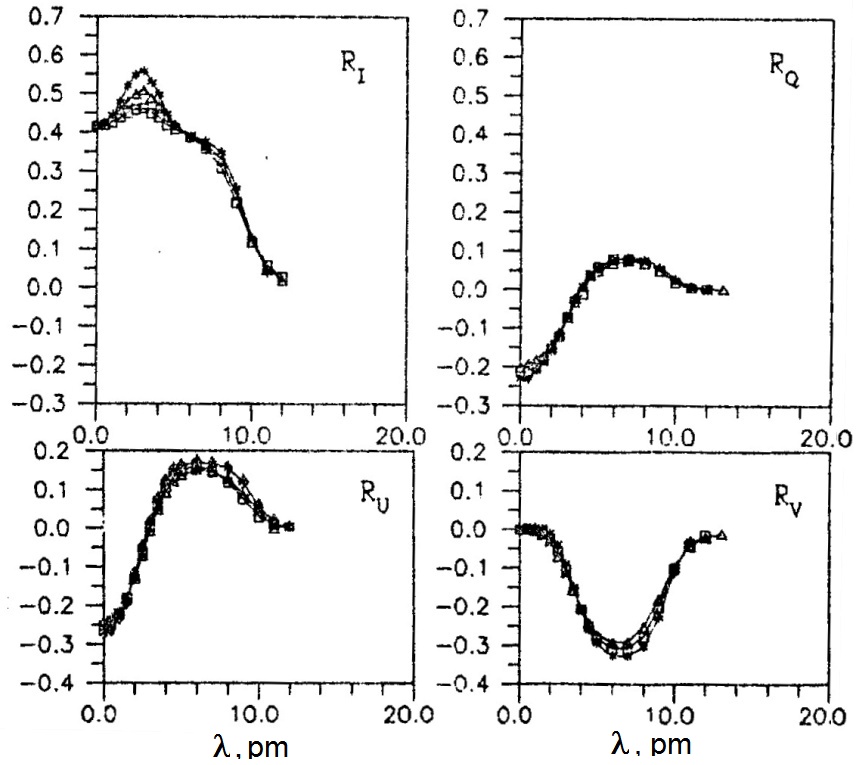}
    }
  \caption {\small\label{11f}
    Stokes profiles of spectral lines with the adjustment factor of the damping constant $E=0$ ($\ast$); 5 ($\square$); 10 ($\triangle$) calculated in the presence of  the magnetic field with $H=1500$~G, $\gamma=55^\circ$, and  $\varphi=30^\circ$.
}
 \centerline{
\includegraphics [scale=0.7]{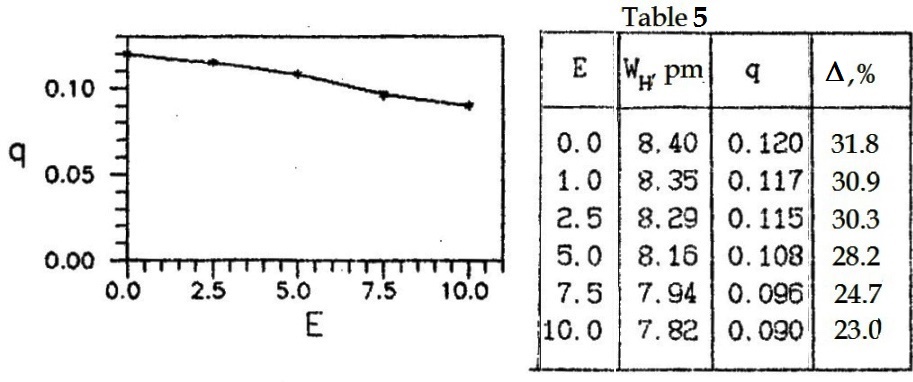}
     }
 \caption {\small\label{12f}
Dependence of the magnetic strengthening $q$ on the adjustment factor  $E$.
\vspace {0.5cm}
\newline Table 5.  The magnetic strengthening $q$ and $\Delta$ of the lines  with the different $E$.}
\end{figure}

\textbf{Lande factor.}  It is natural to expect significant changes in the shape of the Stokes profiles with increasing Lande factor $g_{ef}$.  Figure \ref{7f} shows the profiles for different values of $g_{ef}$. As the Lande factor  increases, the $R_I$ profile becomes wider, their central depth decreases, and their shape becomes more complex. The fraction of polarized radiation grows and the extremums of the profiles $R_Q,~ R_U,~ R_V$ increase. The magnetic strengthening $\Delta$ increases from 3 to 26\% when the Lande factor is increased from 0.5 to 2 and becomes 31\% when $g_{ef}$ reaches a maximum value of 3 (Fig. \ref{8f}, Table 4). It follows that lines with $g_{ef} > 2$ are the most sensitive to magnetic field.

\textbf{Damping constant.} We increased the value of the damping constant $\Gamma= E \gamma_{WdW}$ by changing the correction factor E from 0 to 10.  As can be seen the shape of the Stokes profiles (fig. \ref{11f}) changes only slightly. The magnetic strengthening  $q$ decreases from 32\% to 23\% (Figure \ref{12f}, Table 5) in moderate iron lines.  This seems to indicate that the damping constant affects the profiles much less in lines with low excitation potential  than in those with high $EP$. Note that all lines used have $EP=0.12$ eV. Therefore, we can assume that the dependence of the magnetic strengthening on the damping constant will still be stronger for lines with high excitation potentials.

\section{Influence of atmospheric parameters on Stokes profiles}

Here we consider the influence of the atmospheric model, which specifies the distribution of temperature, pressure, density and chemical composition of the atmosphere.   In addition, we consider the effects of micro-macroturbulent velocities, macroscopic radial velocity, and magnetic field strength, angle of inclination and azimuth.

\begin{figure}
 \centerline{
\includegraphics [scale=0.7]{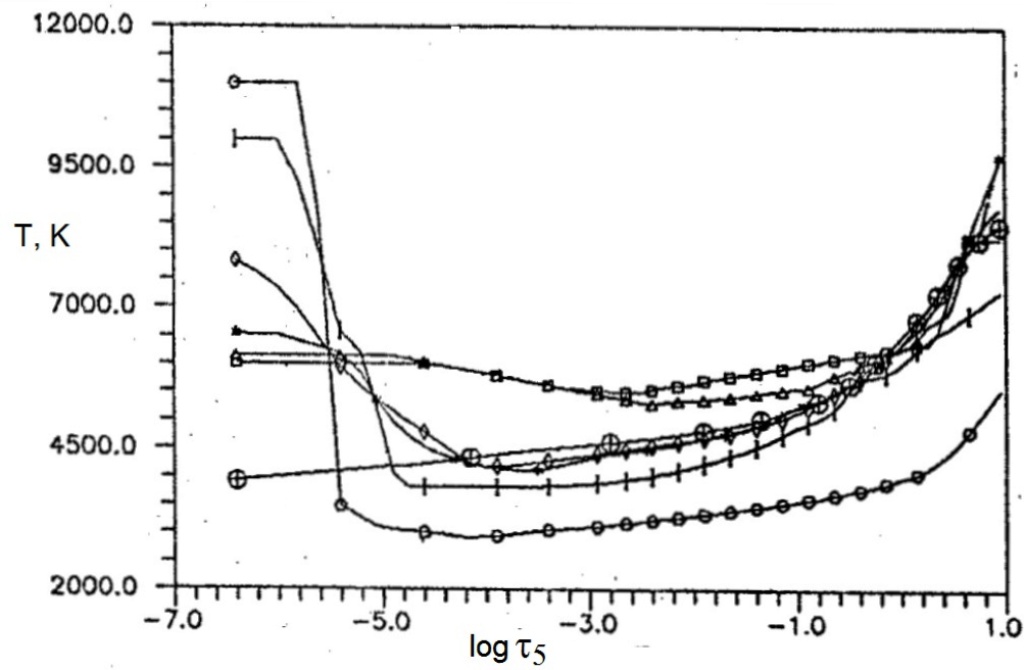}
    }
  \caption {\small\label{13f}
   Temperature distribution of models of the quiet atmosphere of the Sun: HSRA($\lozenge$), VAL($\ast$), HOLMU($\oplus$), the magnetic spot: OBRIDKO1($\circ$), OBRIDKO2(I), and magnetic flux tubes: WALTON1($\triangle$), WALTON2 ($\Box$).
}
\end{figure}

\begin{figure}
 \centerline{
\includegraphics [scale=0.8]{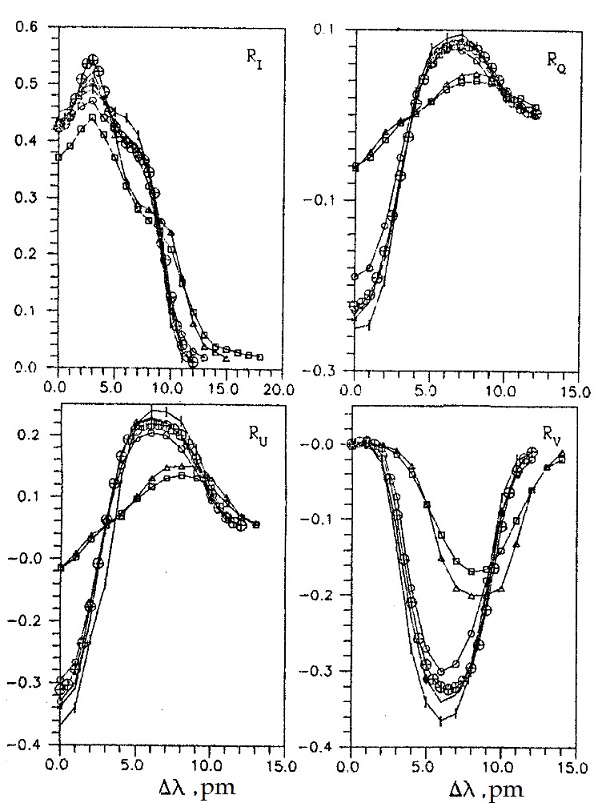}
    }
  \caption {\small\label{15f}
 Stokes profiles calculated in the presence of  the magnetic field with $H=1500$~G, $\gamma=55^\circ$,  $\varphi=30^\circ$ and models of the quiet  Sun: HSRA($\lozenge$), VAL($\ast$), HOLMU($\oplus$),  magnetic spot: OBRIDKO1($\circ$), OBRIDKO2(I), and  flux tube: WALTON1($\triangle$), WALTON2 ($\Box$).
}
\centerline{
\includegraphics [scale=0.6]{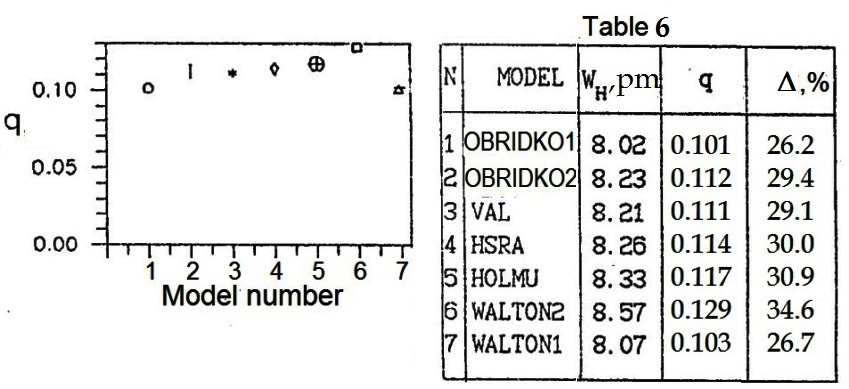}
     }
 \caption {\small\label{14f}
Dependence of the magnetic strengthening $q$ on  atmospheric model marked with a specific number. The symbols are the same as Fig. \ref{15f}.
\vspace {0.3cm}
\newline Table 6.  The magnetic strengthening $q$ and $\Delta$ of the line calculated with different atmospheric models. $N$ is model number.  }
\end{figure}

\textbf{Atmosphere model.} To investigate the effect of temperature, we calculated the profiles with different  models. Figure \ref{13f} shows the temperature as a function of optical depth ($\log \tau_5$) in the following empirical models: a two-component spot model with a cold (OBRIDCO1) and hot (OBRIDCO2) component \cite{8}; two magnetic flux tube models (WALTON1 and WALTON2) \cite{12}: quiet photosphere models HOLMU \cite{7}, VAL \cite{11}, HSRA \cite{5}. Naturally, the temperature distribution has a significant effect on the shape of the Stokes profiles. This can be clearly seen in Fig. \ref{15f} when comparing the profiles calculated with the spot and magnetic  flux tube models, although the magnetic field is the same in these models.  The Stokes profiles calculated with lower temperatures (OBRIDCO1, 2) are similar to those calculated with quiet photosphere models (HOLMU, VAL, HSRA), while the profiles calculated with higher temperatures (WALTON1, 2) are wider and less deep. While the magnetic strengthening $\Delta$ varies slightly between 26\% and 30\% in the lines calculated with the coldest and hottest model (Figure \ref{14f}; Table 6).

\begin{figure}
 \centerline{
\includegraphics [scale=0.8]{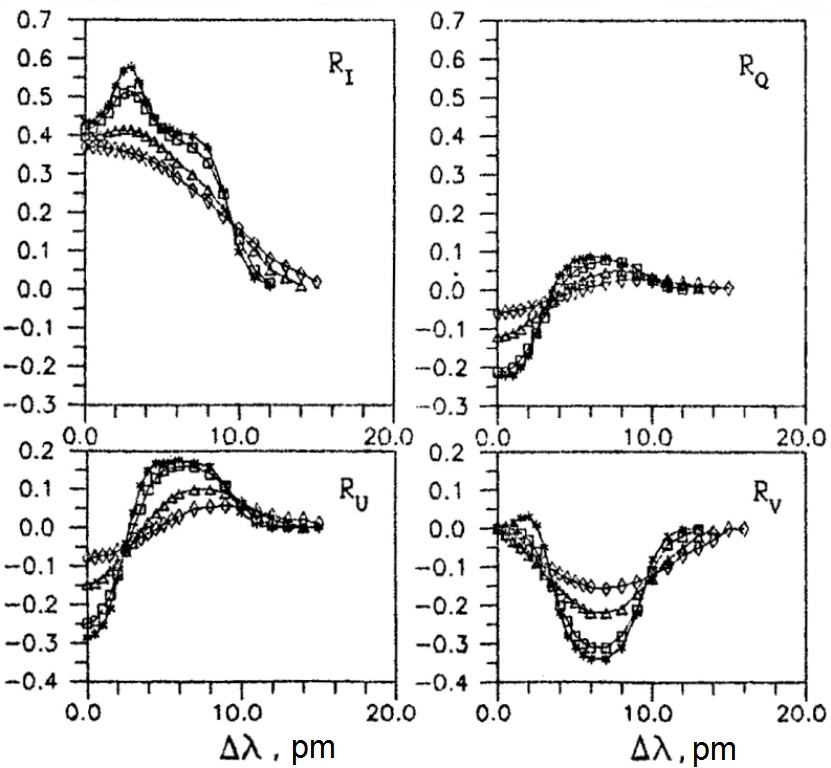}
    }
  \caption {\small\label{9f}
   Stokes profiles calculated with the microturbulent velocity $V_{mic} =0.5$   ($\ast$), 1 ($\Box$),  2 ($\bigtriangleup$),  3 ($\lozenge$) km/s in the presence of  the magnetic field with $H=1500$~G, $\gamma=55^\circ$, and  $\varphi=30^\circ$.
}
 \centerline{
\includegraphics [scale=0.7]{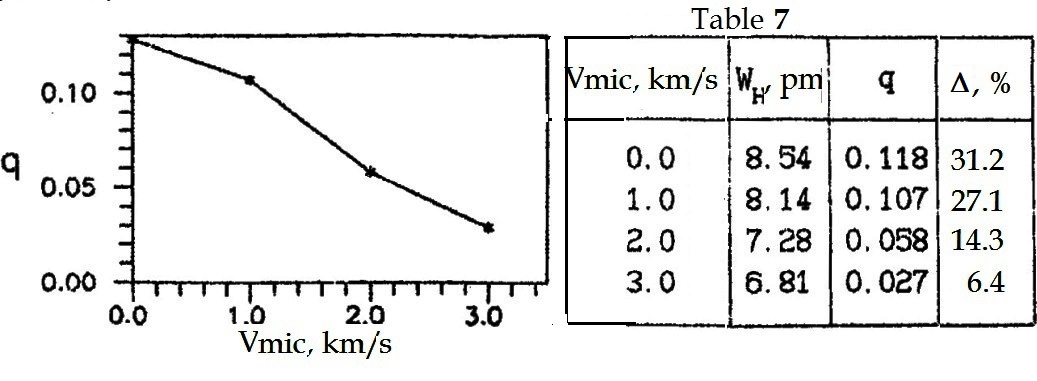}
     }
 \caption {\small\label{10f}
Dependence of the magnetic strengthening $q$ on  microturbulent velocity $V_{mic}$.
\vspace {0.3cm}
\newline Table 7.  The magnetic strengthening $q$ and $\Delta$ of the lines calculated with the different velocities $V_{mic}$. }
\end{figure}

\textbf{Microturbulent velocity.}   According to (5), microturbulence reduces the magnetic broadening of the line, therefore the influence of the magnetic field on the shape of the profiles becomes negligible in atmospheres with high microturbulence velocities.  The higher the microturbulence velocity, the smoother the Stokes profiles (Fig. \ref{9f}) and the less   the equivalent width.  The magnetic strengthening decreases from 31\% to 6\% as microturbulence increases from 0 to 3 km/s (Fig. \ref{10f}, Table 7).

\begin{figure}
 \centerline{
\includegraphics [scale=0.70]{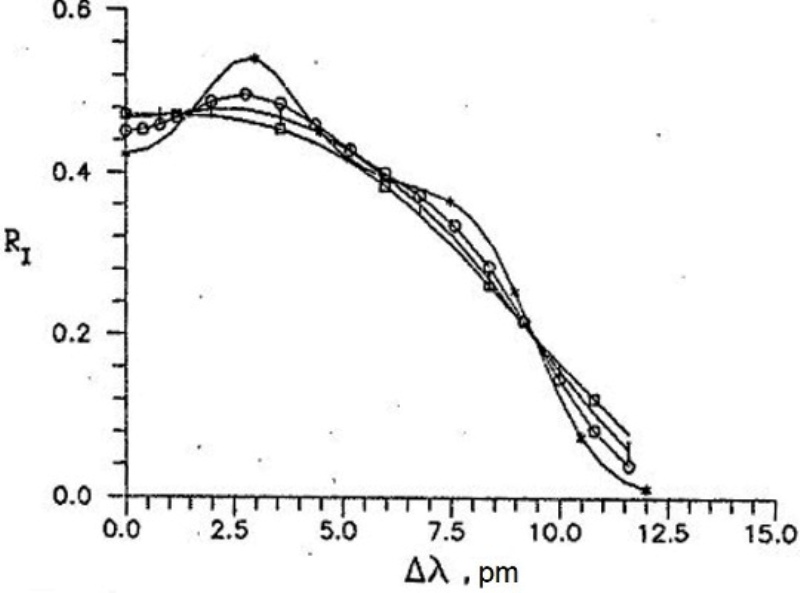}
    }
  \caption {\small\label{25f}
  Stokes profiles calculated with macroturbulent velocities $V_{mac} = 0$ ($\ast$); 1 (o); 1.5 (I); 2 km/s ($\square$) in the presence of  the magnetic field with $H=1500$~G, $\gamma=55^\circ$, and  $\varphi=30^\circ$.
}
\end{figure}

\textbf{Macroturbulence.}  The effect of classical macroturbulence is shown in Fig. \ref{25f}. Macroturbulence also flattens the profile shape, but does not change the equivalent line width. Therefore, the magnetic strengthening is zero.  As can be seen from Fig.  \ref{25f} the characteristic signs of magnetic field influence on the $R_I$ profile disappear at $V_{mac}= 2$ km/s.   The action of macroturbulent velocities on the profiles complicates the recovery of the magnetic field vector using spectral analysis.

\begin{figure}
 \centerline{
\includegraphics [scale=0.5]{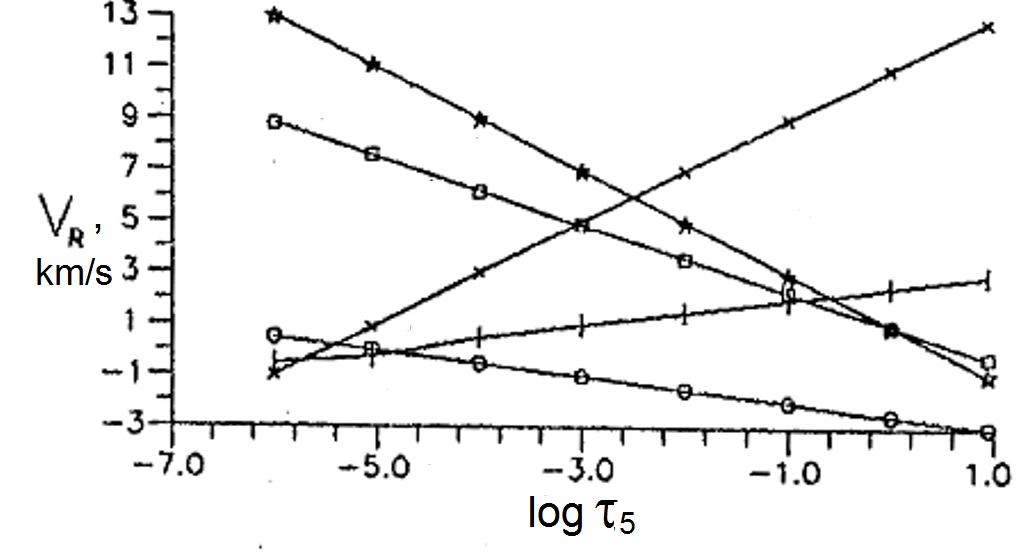}
    }
  \caption {\small\label{26f}
 Dependencies of radial macroscopic velocity with optical depth $V^{rad}$ = $-2 \log \tau_5 +1 $ ($\ast$),  $-1.3 \log \tau_5 +1$ ($\square$),  $2.0 \log \tau_5 +11$ ($\times$), $0.5\log \tau_5 +2.5$ (I),  $-0.5 \log \tau_5 -2.5$ (o).
}
 \centerline{
\includegraphics [scale=0.8]{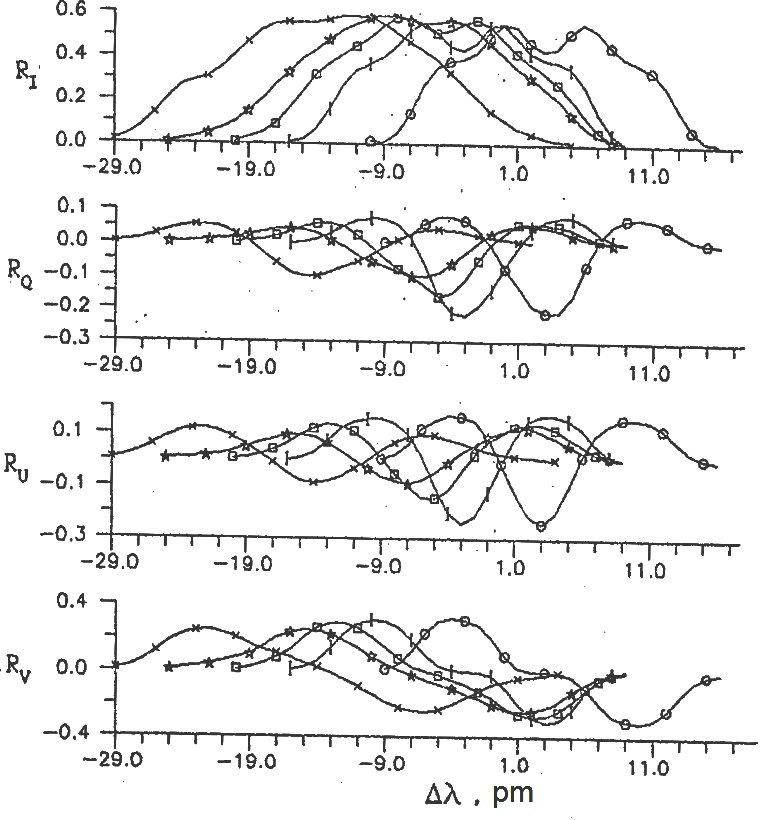}
    }
  \caption {\small\label{27f}
  Stokes profiles calculated in the presence of  the magnetic field with $H=1500$~G, $\gamma=55^\circ$, and  $\varphi=30^\circ$ with different dependencies of radial macroscopic velocities $V^{rad}$, respectively Fig. \ref{26f}.
}
\end{figure}

\textbf{Macroscopic radial velocity}. If the radial velocity $V^{rad}$ is constant with height in the formation region  of the line, then the spectral line is shifted along the wavelength without changing the  profile shape. In real atmospheres, $V^{rad}$ varies with height, so the shifts of different parts of the profile will be different. As a result, the line profile becomes asymmetric. We adopted some $V^{rad}$ dependencies with large gradients to demonstrate clearly the results of this effect on Stokes profiles (Fig. \ref{26f}). Figure \ref{27f} shows how much the shape of the Stokes profiles has changed as a result of the adopted radial velocity gradients. The greater the velocity amplitude and the greater the gradient, the more asymmetrical the shape of the profiles becomes. It seems possible to achieve a good agreement with observations by performing an analysis of the Stokes profiles calculated with the velocity field gradient and the magnetic field gradient, involving the response functions. Although, undoubtedly, this is a  complicated task.


\begin{figure}
 \centerline{
\includegraphics [scale=0.8]{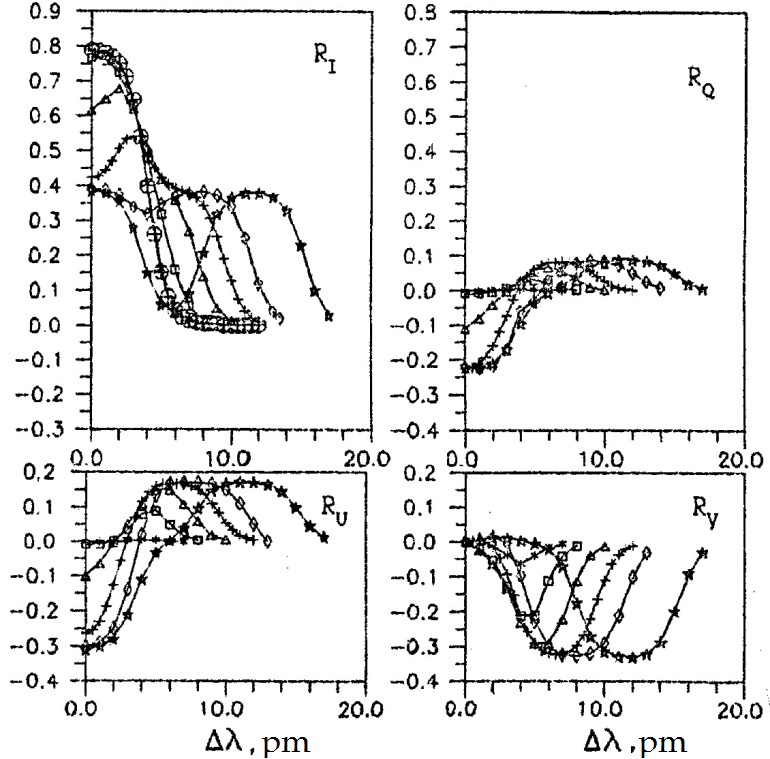}
    }
  \caption {\small\label{17f}
 Stokes profiles calculated in the presence of  the magnetic field with  the strength $H= 0$    ($\ast$),  100 (o), 500 ($\square$), 1000 ($\triangle$), 1500 (+), 2000 ($\lozenge$), 3000 G ($\star$), and  $\gamma=55^\circ$,   $\varphi=30^\circ$.
}
 \centerline{
\includegraphics [scale=0.7]{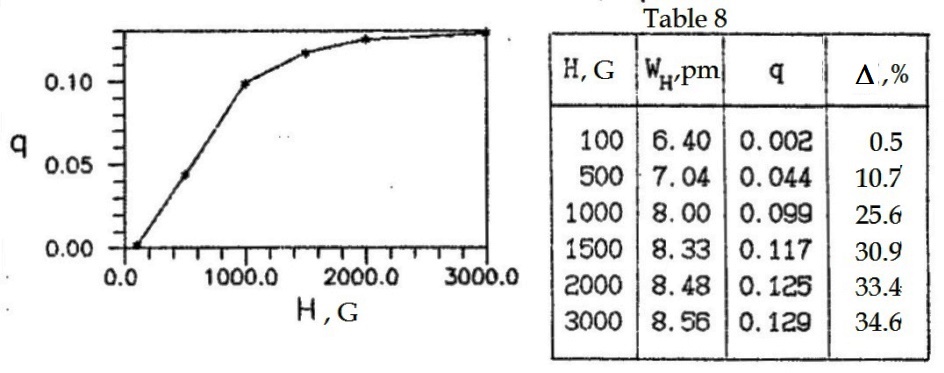}
    }
  \caption {\small\label{16f}
 Dependence of the magnetic strengthening $q$  on  the magnetic field strength  $H$.
\vspace {0.3cm}
\newline Table 8.  The magnetic strengthening $q$, $\Delta$ of the lines with the different $H$.  }
\end{figure}

\textbf{Magnetic field strength.} The higher the magnetic field strength, the more the $R_I$ profile (5) expands, until a complete separation of the splitting components occurs (Fig. \ref{17f}). The shape of the $R_I$ profiles is particularly sensitive to changes in magnetic field strength. The second peak becomes noticeable on the wing of the $R_I$ profile beginning at $H=1000$~G. A complete splitting of the components occurs at $H=3000$~G and the entire profile has three peaks, as seen in Fig. \ref{17f}.   The magnetic strengthening $q(H)$ increases sharply as the magnetic field strength increases from 100 to 1000~G (Fig. \ref{16f}). The increase then slows down and remains almost constant with further increases in $H$. The percentage change in equivalent width increases from 10 to 35\% as the field strength increases from 100 to 3000 G (Table 8).

\begin{figure}
 \centerline{
\includegraphics [scale=0.8]{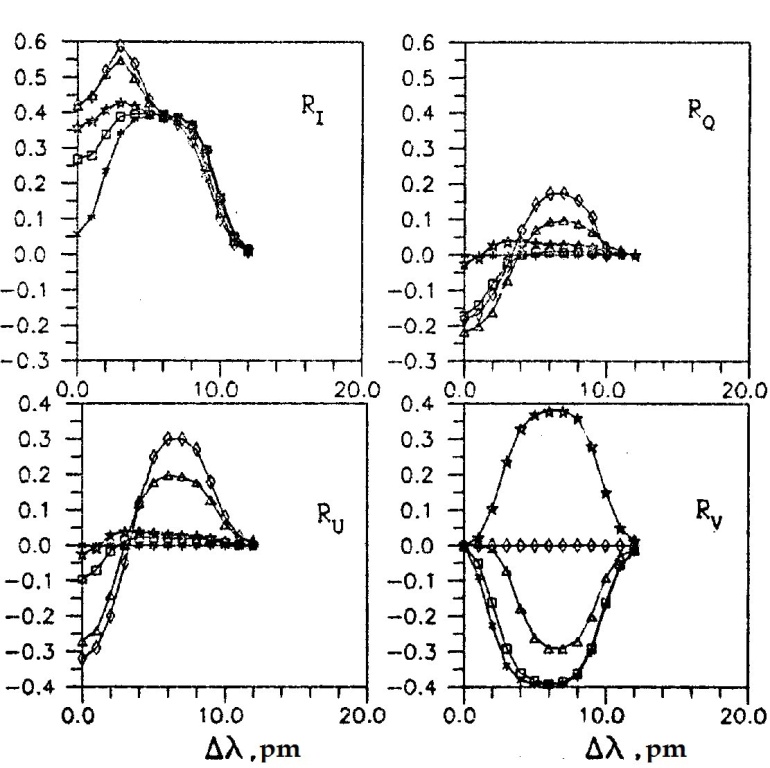}
    }
  \caption {\small\label{18f}
   Stokes profiles calculated in the presence of  the magnetic field with $H=1500$~G, $\gamma= 0^\circ$ ($\ast$); 20$^\circ$ ($\square$); 60$^\circ$ ($\triangle$); 90$^\circ$ ($\lozenge$); 150$^\circ$ ($\star$), $\varphi=30^\circ$.
}
\end{figure}

\textbf{The tilt angle} of the magnetic field vector also strongly changes the shape of the Stokes profiles. Increasing the slope angle $\gamma$ from 0 to 90$^\circ$ affects the contribution of the $\pi$-component and the shape of the $R_I$ profile changes significantly (Figure \ref{18f}). The peaks of $R_Q,~ R_U$ profiles become large and the peak $R_V$ of the profile decreases to 0. A further increase in the tilt angle up to 180$^\circ$ (the field is directed away from the observer and the transverse field becomes longitudinal) results in a change in the $R_V$ sign. The following equations take place:
\begin{eqnarray}
  R_I(\gamma)&=& R_I(180^\circ-\gamma);\nonumber\\
   R_Q(\gamma)&=& R_Q(180^\circ-\gamma);\nonumber\\
    R_U(\gamma)&=& R_U(180^\circ-\gamma);\\
     R_V(\gamma)&=& -R_V(180^\circ-\gamma).\nonumber
\end{eqnarray}
It follows that the profile $R_V$ has a characteristic property. The value of $R_V$ will always be negative if the magnetic field is directed towards the observer ($\gamma= 0$--90$^\circ$ and 270--360$^\circ$).   If the magnetic field is directed away from the observer ($\gamma= 90$--270$^\circ$), then $R_V$ is positive.
\begin{figure}
 \centerline{
\includegraphics [scale=0.7]{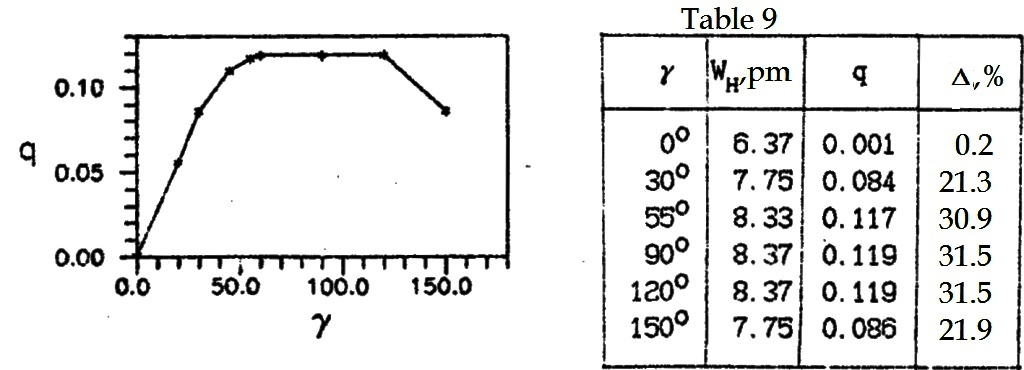}
    }
  \caption {\small\label{19f}
   Dependence of the magnetic strengthening $q$  on the angle of inclination $\gamma$.
\vspace {0.3cm}
\newline Table 9.  The magnetic strengthening $q$, $\Delta$ of the lines  with the different   $\gamma$.  }
\end{figure}
 \begin{figure}
 \centerline{
\includegraphics [scale=0.8]{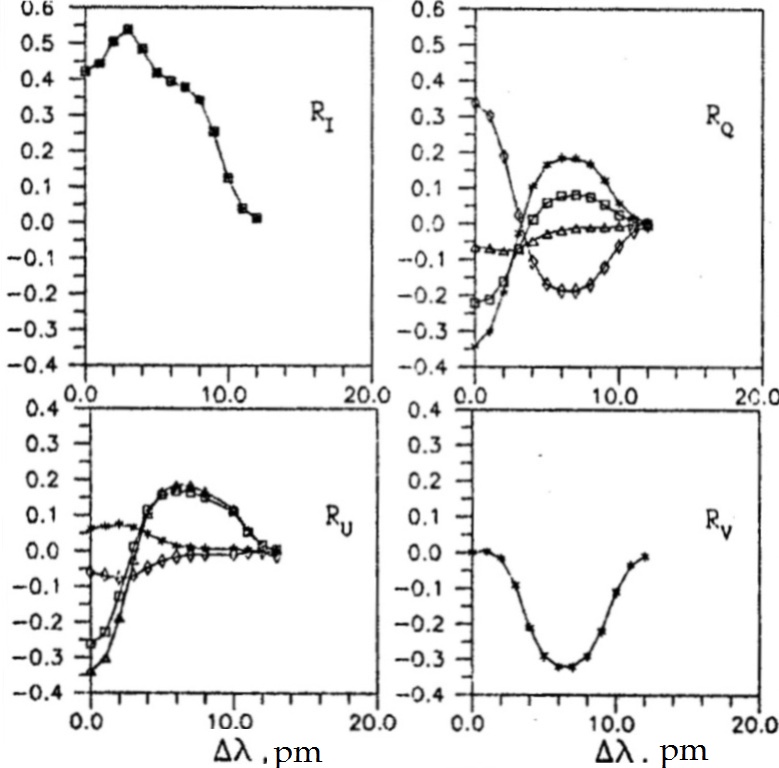}
    }
  \caption {\small\label{20f}
   Stokes profiles calculated in the presence of  the magnetic field with $H=1500$~G, $\gamma=55^\circ$, and $\varphi= 0^\circ$ ($\ast$); 30$^\circ$ ($\square$); 45$^\circ$ ($\triangle$); 90$^\circ$ ($\lozenge$).
}
\end{figure}
 \begin{figure}
 \centerline{
\includegraphics [scale=0.85]{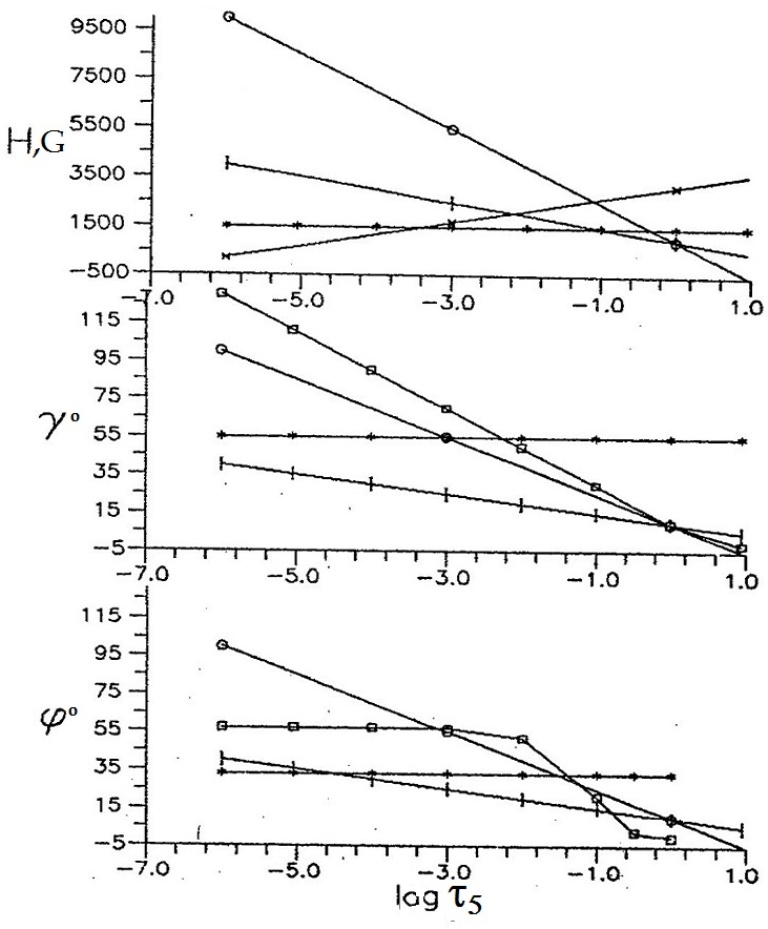}
    }
  \caption {\small\label{21f}
  Dependencies of the magnetic field parameters on optical depth.
   $H=(-1.5 \log \tau_5+1)*1000$ (o), $(-0.5 \log \tau_5+1)*1000$ (I), $(0.5 \log \tau_5+3.2)*1000$ ($\times$), 1500 G ($\ast$). $\gamma =(-1.5 \log \tau_5+1)*10$ (o), $(-0.5 \log \tau_5+1)*10$ (I),$(-2.0 \log \tau_5+1)*10$ ($\square$), $55^\circ$ ($\ast$). $\varphi= (-1.5$, $\log \tau_5+1)*10$ (o), $(-0.5 \log \tau_5+1)*10$ (I), $30^\circ$ ($\ast$),  $\exp (-10\tau_5)$ rad  ($\square$).
}
\end{figure}
 \begin{figure}
 \centerline{
\includegraphics [scale=0.7]{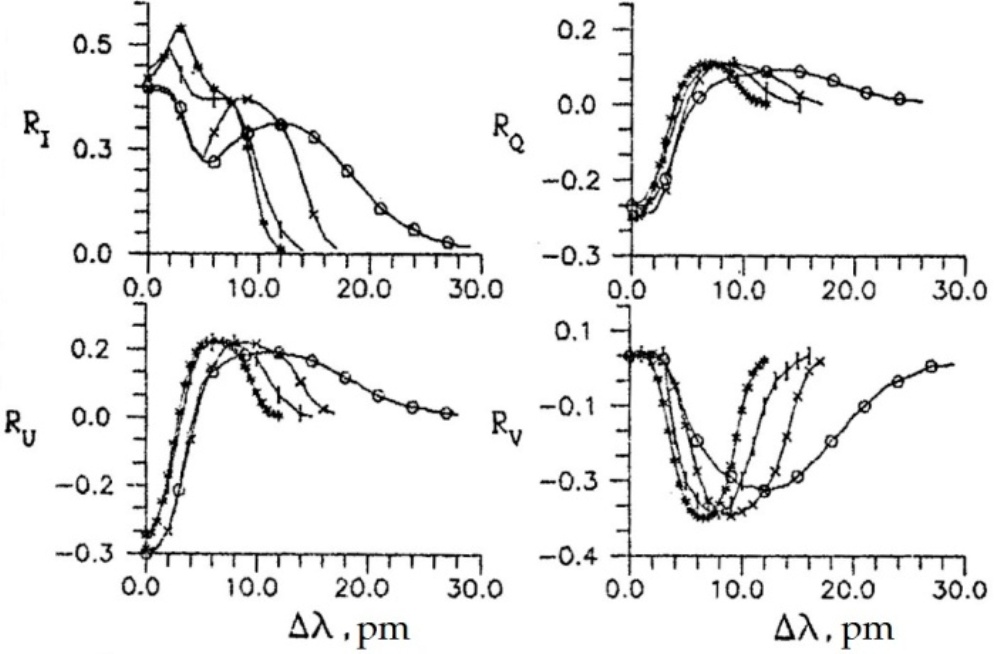}
    }
  \caption {\small\label{22f}
    The Stokes profiles calculated in the presence of  the magnetic field with the different dependencies of the magnetic field strength on optical depth, respectively Fig. \ref{21f}, and  $\gamma=55^\circ$, and  $\varphi=30^\circ$.
}
 \centerline{
\includegraphics [scale=0.7]{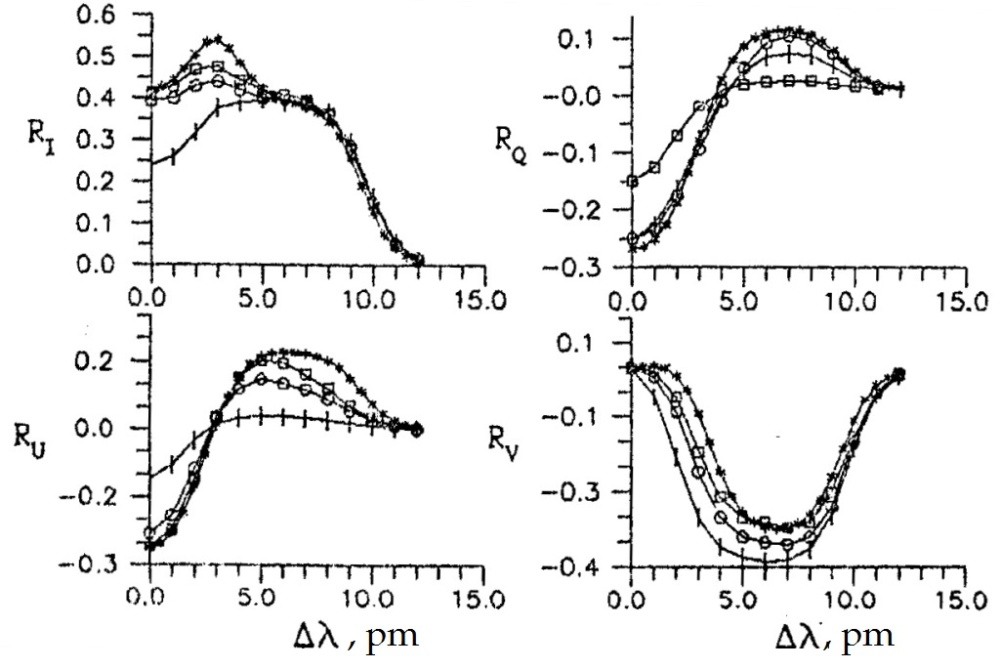}
    }
  \caption {\small\label{23f}
  The Stokes profiles calculated in the presence of  the magnetic field with $H=1500$~G, the different dependencies of the magnetic field tilt angle on the optical depth, respectively Fig. \ref{21f},  and $\varphi=30^\circ$.
}
\end{figure}
 \begin{figure}

 \centerline{
\includegraphics [scale=0.7]{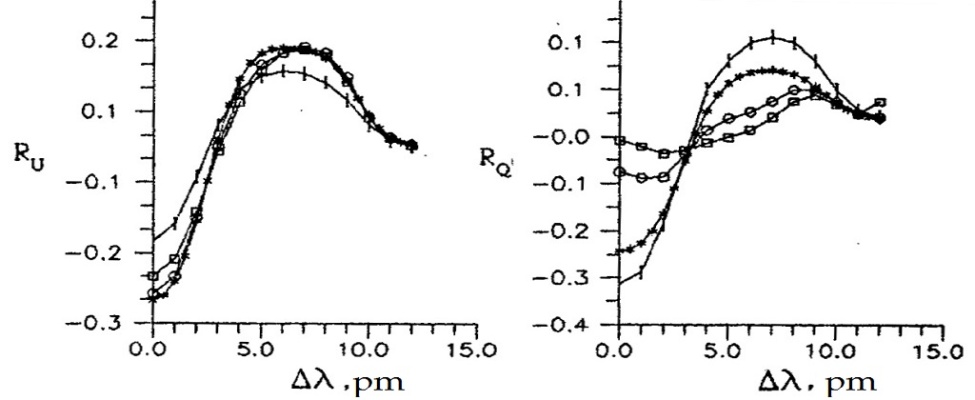}
    }
  \caption {\small\label{24f}
  The Stokes profiles calculated in the presence of  the magnetic field with $H=1500$~G, $\gamma=55^\circ$, and the different dependencies of the magnetic field azimuth on the optical depth, respectively Fig. \ref{21f}.
}
\centerline{
\includegraphics [scale=0.7]{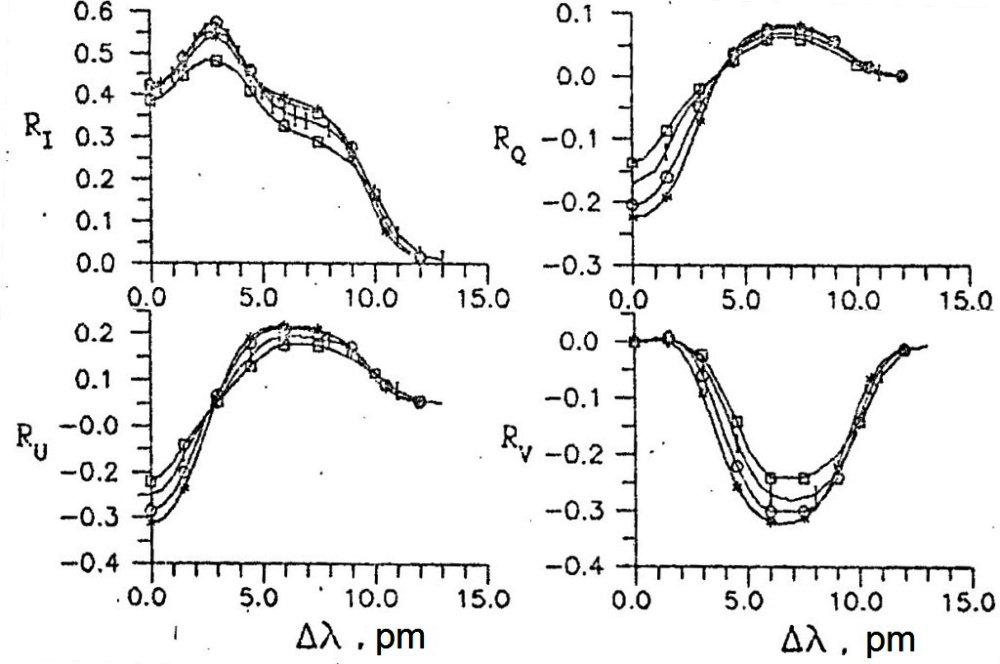}
    }
  \caption {\small\label{28f}
  Stokes profiles calculated in the presence of  the magnetic field with $H=1500$~G, $\gamma=55^\circ$, and  $\varphi=30^\circ$ at different positions on the solar disk $\cos \theta =1$ ($\ast$), 0.6 (o), 0.3 (I), 0.1 ($\square$).
}
\end{figure}
The characteristic property of the $R_V$ profiles allows the main magnetic field direction to be determined.  In principle the magnitude of the tilt angle can be refined by simultaneously comparing the calculated $R_I$ and $R_V$ profiles with the observed ones.  As for the equivalent width and magnetic strengthening, they increase when $\gamma$ changes from 0 to 60$^\circ$ and decrease when $\gamma$ changes from 120 to 150$^\circ$ (Fig. \ref{19f}, Table 9).

\textbf{The magnetic field  azimuth} determines the position of the plane in which the magnetic field vector and the line of sight are located. If the azimuth is changed, the magnitude of the angles at which we observe linearly polarized radiation will also change. The azimuth has no effect on the intensity of the emergent radiation in the line, and the $R_I$ and $R_V$
profiles  remain unchanged, while the $R_Q$ and $R_U$ profiles change (Fig.~\ref{20f}).
\clearpage 
The following equations are valid for the azimuth ($\varphi$):
\begin{eqnarray}
  R_Q(\varphi)&=&-R_Q(90+\varphi)= R_Q(180+\varphi),\nonumber\\
  R_U(\varphi)&=&-R_U(90+\varphi)= R_U(180+\varphi).
\end{eqnarray}
\textbf{Inhomogeneity of the magnetic field.} In real conditions of stellar atmospheres, the strength and direction of the magnetic field change with height. Such magnetic fields are called inhomogeneous. To demonstrate the response of Stokes profiles to magnetic field inhomogeneity, we took three arbitrary height dependencies for strength, tilt angle and azimuth  are presented in  Fig. \ref{21f}.  The profiles calculated according to these dependencies are shown in Fig. \ref{22f}--\ref{24f}. Their shape is sensitive to the field strength gradient in the absorption line formation region. The wider the area of line formation and the larger the gradient, the stronger the effect. It should be emphasized that the width of the line formation region also depends on the magnetic field strength. It is therefore necessary to calculate the contribution  functions  and response functions for the Stokes profiles when we determine the magnetic field parameters from the observed spectrum of polarized radiation.

\textbf{The position on the solar disk} is determined by $\cos \theta$, where $\theta$ is the angle between the line of sight and the normal to the solar surface.  The spectral lines observed at the limb of the solar disk become narrower and shallower than those in the centre of the disk. Figure \ref{28f} shows the Stokes profiles calculated for $\cos \theta = 1.0$, 0.6, 0.3 and 0.1 with a standard value of oscillator strengths $\log gf$.

\section{Anomalous  dispersion}

\begin{figure}
 \centerline{
\includegraphics [scale=1.05]{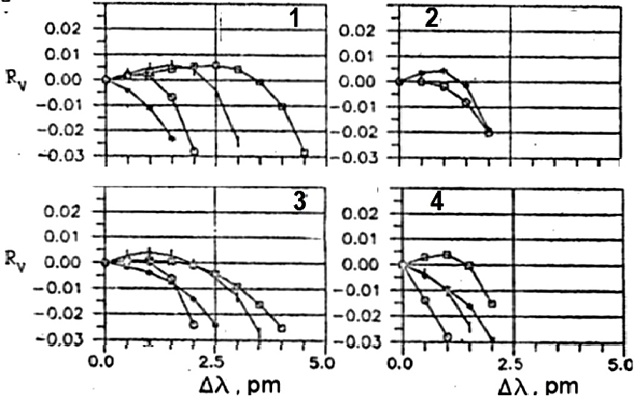}
    }
  \caption {\small\label{29af}
 Anomalous dispersion effect in the central part of the $R_V$ profiles calculated  in the presence of  the magnetic field with $H=1500$~G, $\gamma=55^\circ$, $\varphi=30^\circ$ with the following parameters. \textbf{1} -- $\lambda = 400$ ($\ast$), 500 (o), 600 (I), 700 nm ($\square$); \textbf{2} --  $EP = 0$ ($\ast$), 4 eV (o); \textbf{3} --  $\log gf = 0.5$ ($\ast$), 1.5 (o), 2.0 (I), 3 ($\square$); \textbf{4} --  $g_{ef}= 0.5$ ($\ast$), 1.5 (o), 2.0 (I), 3.0 ($\square$).
}
\end{figure}

\begin{figure}
 \centerline{
\includegraphics [scale=1.05]{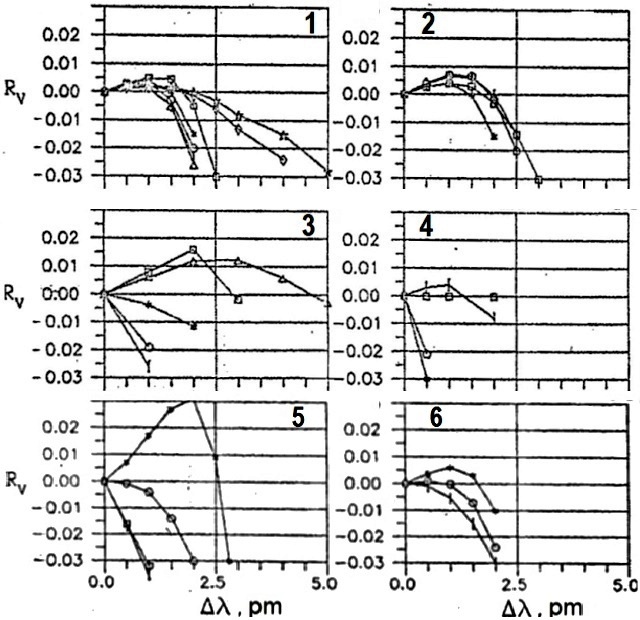}
    }
  \caption {\small\label{30af}
Anomalous dispersion effect in the central part of the $R_V$  profiles calculated  in the presence of  the magnetic field with $H=1500$~G, $\gamma=55^\circ$, and  $\varphi=30^\circ$ with the following parameters. \textbf{1} -- models of  quiet atmosphere of the Sun HSRA ($\ast$), VAL (o), HOLMU (I), models of magnetic spot OBRIDKO1 ($\square$), OBRIAKO2 ($\triangle$); models of magnetic flux tubes WALTON1 ($\star$), WALTON2 ($\lozenge$); \textbf{2} -- positions on the solar disk $\cos \theta = 1$ ($\ast$), 0.6 (o); 0.3 (I); 0.1 ($\square$); \textbf{3} -- magnetic field strength $ H= 100 $ (I), 500 (o), 1500 ($\ast$), 2000 ($\square$), 3000 G ($\triangle$); \textbf{4} -- tilt angle of the magnetic field vector $\gamma = 0^\circ$ ($\ast$), 20$^\circ$ (o), 60$^\circ$ (I), 90$^\circ$ ($\square$); \textbf{5} -- $V_{mic}= 0$ ($\ast$), 1 (o), 2 (I), 3 km/s ($\square$); \textbf{6} -- $E = 0$ ($\ast$), 5 (o), 10 (I).
}
\end{figure}

We also consider the effects of anomalous dispersion, which appear only in the $R_V$ profiles of circularly polarised radiation. In our calculations, the anomalous dispersion was accounted for according to Raczkowski's theory \cite{14,15}. Signs of anomalous dispersion appear in the wings of the $R_V$ profile near its centre.  Here the sign of $R_V$ is reversed with respect to the rest of the wing. The magnitude of the anomalous dispersion in the $R_V$ profile is clearly visible in Fig. \ref{29af} and Fig. \ref{30af}. As can be seen, the effect of anomalous dispersion increases with increasing wavelength, magnetic field strength and decreases with increasing excitation potential, attenuation constant and microturbulence. It also increases at the solar limb. There is no clear dependence on oscillator strength (or equivalent width), effective Lande factor, temperature, angle of inclination of the magnetic field. In general this effect is small, but nevertheless becomes clearly visible in atmospheres with very low microturbulent velocities ($V_{mic}<0.5$) and with strong magnetic fields ($H>2000 G$).
\section{Conclusions}

In this work we have considered the change of spectral lines in the photosphere of the Sun in the presence of a magnetic field.  Analysis of the results allows us to draw the following conclusions.

 The shape of the Stokes profiles depends mainly on line strength, wavelength, and the Lande factor.  The best lines to study magnetic fields in stellar atmospheres are moderate lines ($W=7$--11 pm) with the longest possible wavelength and a large Lande factor.

Regarding the parameters of the medium in which the spectral lines are formed, the magnetic field strength, the inclination angle of the magnetic field vector, and the velocity field have the greatest influence on the Stokes profiles.  The magnetic field strength significantly changes the shape of the total radiation profile and introduces characteristic features that become more pronounced as the strength increases. At the same time, the polarized radiation profiles become broader and stronger. The tilt angle affects the polarised radiation profiles to a greater extent. Micro- and macroturbulent velocities smooth out the features of the profiles associated with the magnetic field. Therefore, the shape of the observed Stokes profiles in the atmospheres of stars with microturbulence greater than 3~km/s will not correspond to the real magnetic fields. Macroscopic radial velocity shifts the line as a whole. Magnetic and velocity field gradients in the line formation region add asymmetry to the Stokes profiles.

The spectral line always becomes wider under the influence of the magnetic field, and its equivalent width increases.  In other words, there is a magnetic strengthening of the line in the magnetic field. According to our results, the magnetic strengthening little dependent on  wavelength, excitation potential, damping constant, atmospheric model and magnetic field azimuth.  It depends mainly on the Lande factor, the magnetic field strength and the inclination angle. The larger the Lande factor and the magnetic field strength, the larger the magnetic strengthening. The greatest strengthening occurs when the angle of inclination is 55--120. The magnetic strengthening grows up to a certain limit, after which the saturation effect sets in and the growth stops.   Moderate lines ($W=7$--11 pm ) are the most sensitive to magnetic field variations. The percentage strengthening of such lines is maximum and therefore they can be successfully used to investigate the structure of the magnetic field. For strong lines ($W=20$ pm), the percentage magnetic strengthening is 2.7 times smaller, and for weak lines ($W=2$ pm) 10 times smaller than for moderate lines.

The profiles of circularly polarized radiation are sensitive to anomalous dispersion which appears as a small projection with opposite sign near the centre of the profile on both wings and is proportional to the magnetic strengthening. The magnitude of this effect increases slightly with wavelength and equivalent width, and increases more markedly with magnetic field strength. The anomalous dispersion decreases sharply with increasing amplitude of microturbulent velocity and is not observed at velocities of 1 km/s.

Analysis of the four observed and computed Stokes profiles will provide additional information about the magnetic field structure. In our view, the inclusion of Stokes profiles in spectral analysis of the atmosphere is necessary.  The promise of Stokes profile studies should not be in doubt. However, our analysis also shows that comparing calculated and observed Stokes profiles may not give unambiguous results. New methods for analysis of Stokes profiles need to be developed.

In conclusion, the author hopes that on the basis of the results presented here it will be possible to: 1) make a correct choice of magnetically active lines for modelling structural irregularities of the solar photosphere, such as flares, spots and other formations, for which models are currently being actively developed; 2) determine initial values of input parameters when recovering the magnetic field vector by analysing spectropolarimetric line profiles with diagnostic methods.


\end{document}